\documentclass[pra,aps,twocolumn,showpacs,superscriptaddress,floatfix,amsmath,footinbib,amssymb]{revtex4-1}

\usepackage{graphicx}
\usepackage{mathrsfs}
\usepackage{color,xspace}
\usepackage[colorlinks=true]{hyperref}
\usepackage{units}

\definecolor{ForestGreen}{rgb}{0.3,0.6,0.3}
\definecolor{MidnightBlue}{rgb}{0.023,0.375,0.559}
\hypersetup{linkcolor=MidnightBlue,citecolor=MidnightBlue,urlcolor=MidnightBlue}

\newcommand{\proj}[2]{\left| {#1} \right\rangle\!\left\langle {#2} \right|}
\newcommand{\tr}{\operatorname{tr}}

\newcommand{\ra}{\right\rangle}
\newcommand{\ket}[1]{\left| #1 \ra}

\newcommand{\modsq}[1]{\left| #1 \right|^2}
\newcommand{\lb}[1]{\log_2\left( #1 \right)}
\newcommand{\AR}{\overline{\text{R}}}
\newcommand{\ro}{r_{\omega}}
\newcommand{\eq}[1]{(\ref{#1})}
\newcommand{\qr}{q_\mathrm{R}}

\begin{document}

\title{Fundamental limitations to information transfer in accelerated frames}

\author{Eduardo Mart\'in-Mart\'inez}
\affiliation{Institute for Quantum Computing, Department of Physics and Astronomy and Department of Applied Mathematics, University of Waterloo, 200 University
Avenue W, Waterloo, Ontario, Canada, N2L 3G1}
\affiliation{Instituto de F\'{i}sica Fundamental, CSIC, Serrano 113-B, 28006 Madrid, Spain}

\author{Dominic Hosler}
\email{dominichosler@physics.org}
\affiliation{Department of Physics and Astronomy, University of Sheffield, Hicks building, Hounsfield road, Sheffield, S3 7RH, United Kingdom}

\author{Miguel Montero}
\affiliation{Instituto de F\'{i}sica Fundamental, CSIC, Serrano 113-B, 28006 Madrid, Spain}

\begin{abstract}
We study communication between an inertial observer and one of two causally-disconnected counter accelerating observers. We will restrict the quantum channel considering inertial-to-accelerated bipartite classical and quantum communication over different sets of Unruh modes (single-rail or dual-rail encoding).
We find that the coherent information (and therefore, the amount of entanglement that can be generated via state merging protocol) in this strongly restricted channel presents some interesting monogamy properties between the inertial and only one of the accelerated observers if we take a fixed choice of the Unruh mode used in the channel.
The optimization of the controllable parameters is also studied and we find that they deviate from the values usually employed in the literature.
\end{abstract}


\maketitle

\section{Introduction}
\label{sec:introduction}

Physics can be formulated in the language of information theory \cite{Wheeler1989}.
Such a program models all interactions between particles as the transfer of information.
As the physical world is fundamentally based on quantum physics, these interactions can mediate quantum information.

The communication abilities of classical channels have been quantified using Shannon's noisy coding theorem \cite{Shannon1949}.
When considering a quantum description of a communication medium, the specification of a channel code is very different \cite{Wilde2011,Horodecki2005}, and different measures quantify achievable rates for communication \cite{DiVincenzo1998a,Devetak2004,Cliche2010,Bradler2010a,Hsieh2008,Fawzi2011}, even in the case of classical transmission over quantum channels \cite{Shor2004,Schumacher1997,Fawzi2012,Holevo1996}.
In relativistic settings there is noise caused by noninertial motion, or the curvature of space time.
For accelerated frames and black holes this noise is due to the Unruh-Hawking effect \cite{Takagi1986,Davies1975}.

In the context of relativistic quantum information, field entanglement in noninertial settings has been thoroughly studied \cite{Fuentes-Schuller2005,Alsing2006,Martin-Martinez2009,Martin-Martinez2010c,Martin-Martinez2010b,Leon2009,Martin-Martinez2010a,Martin-Martinez2012b}.
When entanglement is used in conjunction with classical and quantum communication, it can be useful to determine achievable rate triples, which can lead to trade-offs between classical communication, quantum communication, and entanglement consumption.
The Unruh channel's rate triples have been studied within the single mode approximation \cite{Wilde2012a,Bradler2010,Wilde2011b}.
Since the extension of the formalism beyond the single mode approximation (SMA) \cite{Bruschi2010}, field entanglement has been explored by means of entangled states of the inertial vacuum and Unruh modes, i.e., acceleration dependent families of inertially maximally entangled states that depend upon a parameter $r$ which depends itself on acceleration.

In this paper we study the communication between two pairs of observers: an inertial observer, Alice ($A$), and two constantly accelerated complementary observers Rob ($R$) and anti-Rob ($\AR$) moving with opposite accelerations in two causally disconnected regions of the Rindler space time, the same setting as in \cite{Bruschi2010}.

It has been suggested that the single mode approximation is optimal for quantum communication between Alice and Rob \cite{Hosler2012}.
However, there are interesting consequences when one considers a setting of three observers Alice--Rob--anti-Rob.
For those cases the SMA is not sufficient, and  if we want to set up a system in which Alice has a chance of communicating with either of the accelerated observers, Rob or anti-Rob, then we need to move beyond the single mode approximation. If we also want to consider the possibility of Alice communicating with both at the same time, then we need to consider the full formalism of quantum broadcast channels where, as shown in \cite{Dupuis2011}, quantum communication to either party is possible.

This last point is worth stressing. Notice that our setting is genuinely a broadcast scenario (we have one sender and two receivers). Communication strategies have been known for some time in the context of classical communication over classical broadcast channels \cite{Cover1972}, and Dupuis \emph{et al.} have now determined communication strategies for quantum communication over quantum broadcast channels \cite{Dupuis2011}. However, the analysis of this work will be  much more  modest: we will analyze the coherent information along the restricted channel where the choice of Unruh modes is fixed for the communication of Alice with Rob and anti-Rob. Given that standard communication protocols for a broadcast channel are completely inadequate for achieving the true capacity region of such a channel we are, therefore, not claiming that this is the optimal strategy.

The reason for this strong restriction over the channel is that we want to show that, for families of states built from Unruh modes, Alice and different noninertial observers may not be able to generate entanglement with these quantum channels by employing the state merging protocol.
In other words, when preparing the field state in an Unruh mode, Alice will have to choose with whom she wants to generate entanglement, since some of the states that she can prepare will not allow for entanglement generation with some of the noninertial observers.
We will show that this holds when we go beyond the SMA, namely that for each fixed choice of Unruh mode, there is one of the bipartitions for which the coherent information vanishes.
Furthermore, we will see that for the classical channel the SMA is not sufficient even for the Alice-Rob case, since Holevo information within the SMA is not maximal even for the simplest case studied in \cite{Hosler2012}. Instead, one has to take into account the different behaviour of the Holevo information between Alice and the accelerated observers for states beyond the SMA.

\section{Information measures}
\label{sec:informationmeasures}

We would like to quantify the amount of information (classical and quantum) that Alice can communicate to Rob and anti-Rob by means of two standard communication protocols.

The Holevo information \cite{Wilde2011,Holevo1973} quantifies the amount of classical information transferred through a quantum channel and it is defined as
\begin{equation}
  \label{eq:holevoinfo}
  I(A;R)_\sigma = S(\sigma_A)+S(\sigma_R)-S(\sigma_{AR}),
\end{equation}
with respect to the classical quantum state,
\begin{equation}
  \label{eq:classicalquantumstate}
  \sigma_{AR} = \sum_x p_A(x) |x\rangle_A\langle x| \otimes \sigma_{R_x},
\end{equation}
where $x \in \{0,1\}$ and $\sigma_{R_x}$ is the state Rob receives when Alice sends the logical value $x$.

For quantum communication over a quantum channel we require a different measure.
As is commonplace in the literature \cite{Horodecki2005}, we will use the conditional entropy,
\begin{equation}
  \label{eq:CEdefinition}
  S(A|R)_\rho = S(\rho_{AR})-S(\rho_R),
\end{equation}
where
\begin{equation}
  \label{eq:vonneumannentropy}
S\left(\sigma\right) = -\tr\left(\sigma \lb{\sigma}\right) = -\sum_i \left(\lambda_i \lb{\lambda_i}\right)
\end{equation}
is the von Neumann entropy and $\lambda_i$ are the eigenvalues of $\sigma$ \cite{Wilde2011}.

This is interpreted as the amount of quantum information required by Rob for the state merging protocol \cite{Horodecki2006}.

Suppose that Alice and Rob initially share many copies of some bipartite state.
State merging is a protocol whereby classical communication and entanglement are employed to transfer Alice's share of this state so that, at the end of the protocol, Rob possesses the entire state.
Conditional entropy is a measure of how much information Alice is required to send to Rob for this to be possible.

Unlike the classical case, the quantum conditional entropy can be negative, which is interpreted as Rob having an excess of quantum information so that the state merging protocol generates entanglement rather than consumes it.  Hence taking the negative of this conditional entropy gives us the quantum coherent information, measuring quantum correlations,
\begin{equation}
  \label{eq:coherentinfodefinition}
  I(A\rangle R)_\rho =- S(A|R)_\rho.
\end{equation}
This coherent information is the amount of entanglement gained between Alice and Rob by performing the state merging protocol.
This entanglement can be used for communication in the future provided that classical communication is possible, and is related to the quantum channel capacity \cite{Horodecki2005}.

We will analyse here the information flow for two different idealized communication scenarios:
(1) the single-rail channel, which uses a single field mode, representing a logical zero with the vacuum and a logical one with a single excitation and
(2) the dual-rail channel, which uses excitations of two different field modes to represent logical zero and logical one.

We will separately calculate the information measures in the Alice-Rob and Alice--anti-Rob bipartitions.
The study of multiparty broadcast channels with a less naive approach and considering all the subtleties of communication strategies in these kinds of settings (see \cite{Dupuis2011,Cover1972,Savov2011,Guha2007,Yard2011}) is a topic for future research.
Full multiparty communication is forbidden as Rob and anti-Rob are causally disconnected.

\section{Setting}
\label{sec:communicationmethods}
To study communication channel capacities, we must optimize over all parameters and encodings controlled by either Alice or Rob. The calculation of the optimum achievable rates within this restricted channel scenario (single- and dual-rail encodings) is dealt with by means of a numerical optimization over such parameters.

Alice has the freedom to choose which field mode to excite to send a message to Rob.
We will not consider the possibility of Alice or Rob using arbitrary elements of the Fock space, as this would mean an infinite number of optimization parameters. This problem, common when dealing with bosonic quantum channels, is often tacked by imposing a mean-photon-number constraint. However, for our purposes, we will stick to the simpler case of single field modes excited just once.

For the quantum channel case, we will consider two different states of a bosonic field,
\begin{align}
\ket{\psi_S}&=\alpha\ket{00}+\beta\ket{11}, \label{eq:psiS}\\
\ket{\psi_D}&=\alpha\ket{1_+1_+}+\beta\ket{1_-1_-},\label{eq:psiD}
\end{align}
where $\ket{1_{\pm}}$ are respectively zero-mode and one-mode excitations for the dual-rail scenario.  For $\modsq{\alpha}=\modsq{\beta}=\frac{1}{2}$, the state $\ket{\psi_S}$ is a maximally entangled superposition of the Minkowskian vacuum and a pair of non-localized excitations (see Sec.~\ref{sec:transformation}).
They will be created by Alice and then the second will be accessed by the accelerated observers.
This is a kind of state that has been thoroughly studied throughout previous literature \cite{Fuentes-Schuller2005,Alsing2006,Martin-Martinez2009,Martin-Martinez2010c,Martin-Martinez2010b,Leon2009,Martin-Martinez2010a,Bruschi2010,Montero2011b}.
Here we use it to study the single-rail channel.

We use the parameter values,
\begin{equation}
 \label{eq:alphabeta}
 \modsq{\alpha}=\modsq{\beta}=\frac{1}{2},
\end{equation}
throughout most of this article.
In dual-rail encoding the noise is symmetric in both the zero and one modes.
This symmetry strongly suggests that this choice of $\alpha$ and $\beta$ is optimal in the dual-rail case.
We verified this numerically.
For the single-rail case, we also use these values as it provides representative behaviour, and is near the maximum as detailed below.

$\ket{\psi_D}$ is a maximally entangled superposition of two bosonic excitations with a different state for some two-dimensional internal degree of freedom that Alice is able to observe.
This kind of state is studied in the higher spin generalizations of field entanglement analysis formalisms \cite{Montero2011b} where the electromagnetic field case is analysed.
Here we associate one of the internal degrees of freedom with the zero mode and the other with the one mode and use it to study the dual-rail channel.

For the case of classical communication we use a probabilistic mixture of a logical zero or a logical one being sent.
This is represented in the quantum bipartite states
\begin{equation}\label{eq:state2}
\sigma_S=\modsq{\alpha}\proj{00}{00}+\modsq{\beta}\proj{11}{11}
\end{equation}
for the single-rail case and
\begin{equation}\label{eq:state3}
\sigma_D=\modsq{\alpha}\proj{1_+1_+}{1_+1_+}+\modsq{\beta}\proj{1_- 1_-}{1_-1_-}
\end{equation}
for dual-rail communication.

Similar to the quantum communication, we use the values for $\alpha$ and $\beta$ in Eq.~\eqref{eq:alphabeta}.
This is again optimal for the dual-rail case due to symmetry, and near optimal for the single-rail case.
We discuss optimality further in Sec.~\ref{sec:results}.

\section{Transformation between Alice, Rob and anti-Rob's modes}
\label{sec:transformation}
Alice lives in flat space and uses Minkowski modes to describe the field state (a complete set of plane-wave solutions of the field equation in Minkowskian coordinates).
Rob and anti-Rob can together fully describe the field with Rindler modes  (a complete set of plane-wave solutions of the field equation in the corresponding Rindler coordinates).
It is easier to relate these two descriptions through a different complete set of solutions of the field equation from the Minkowskian perspective and that map to single frequency Rindler modes.
These solutions are called Unruh modes and they share a vacuum with Minkowski modes.

We will only consider Unruh modes of a given Rindler frequency $\omega$ as seen by Rob or anti-Rob (who move with proper acceleration $a$).
An arbitrary single frequency Unruh mode for a given acceleration has the form \cite{Bruschi2010}
 \begin{align}C_\omega=q_\text{L}C_{\omega,\text{L}}+\qr C_{\omega,\text{R}}\label{umodes},\end{align}
where $\vert q_\text{L}\vert^2+\vert\qr\vert^2=1$, $\qr \ge q_\text{L}$, and the appropriate expressions for the operators in \eqref{umodes} for the scalar case are,
\begin{eqnarray}\label{bogoboson}
 C_{\omega,\text{R}}&=&\cosh r_\omega\, a_{\omega,\text{I}} - \sinh r_\omega\, a^\dagger_{\omega,\text{II}},\\*
 C_{\omega,\text{L}}&=&\cosh r_\omega\, a_{\omega,\text{II}} - \sinh r_\omega\, a^\dagger_{\omega,\text{I}}, \end{eqnarray}
where $a_\omega,a_\omega^\dagger$ are Rindler particle operators for the scalar field in each region.

Note that what was called in the literature the single mode approximation is just the choice of Unruh modes such that $q_\text{R}=1$. Hence, when we discuss our results ``beyond the single mode approximation'' what we mean is that we study different kinds of Unruh modes with different values of  $q_\text{R}$ just in the same sense as in \cite{Bruschi2010}.

Therefore, the excitations in \eqref{eq:psiS}, \eqref{eq:psiD}, \eqref{eq:state2}, and \eqref{eq:state3} are considered for convenience to be Unruh modes, where $\ket{1}=C^\dagger_\omega\ket{0}$ is the Unruh particle excitation. All these states have an implicit dependence on Rob's acceleration $a$ when expressed in the Rindler basis through a parameter $r_\omega$ defined by $\tanh r_\omega=e^{-\pi c\,\omega/a}$.

Due to the delocalization of Unruh modes they  are arguably not completely measurable, and in the best case scenario they can be only approximately determined (by means of localized measurements).
Also, the Unruh modes, as seen from any inertial observer, behave in a highly oscillatory way near the acceleration horizon.
This makes them bad candidates for physically feasible states.
Finally, we are choosing a different Unruh mode for each acceleration, so as to keep the Unruh to Rindler change of basis always simple.
This simplification is common in the literature.

However, we use them as the states built from these modes allow direct Bogoliubov transformations between Unruh and Rindler modes. This maximises the information transfer of the channels we study providing a limit to the entropic functions. Also, these modes in this setting have been used widely throughout the literature in recent years, so this allows us to compare and discuss with previous works \cite{Bruschi2010}.

\begin{figure}[tb]
  \begin{center}
      \includegraphics[width=85mm]{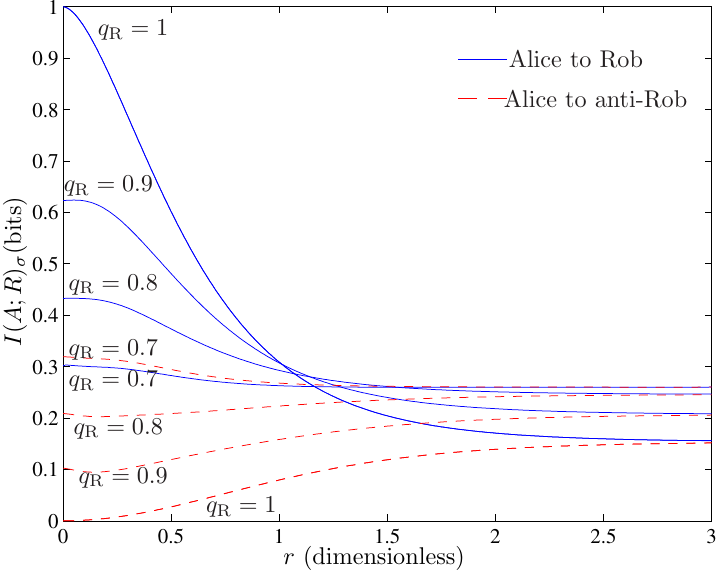}
  \caption[Single rail Holevo information]{(Color online) Holevo information of the single-rail case of classical communication.}  \label{fig:singleMI}
  \end{center}
\end{figure}

\begin{figure}[tb]
  \begin{center}
      \includegraphics[width=85mm]{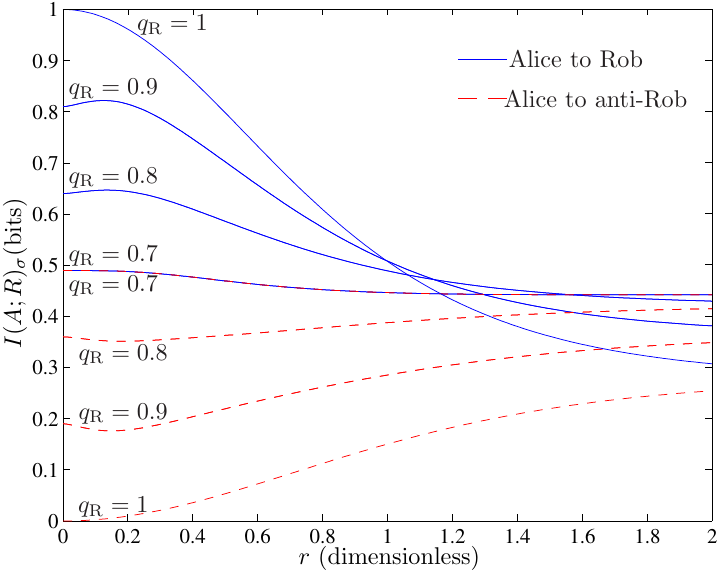}
  \caption[Dual rail Holevo information]{(Color online) Holevo information of the dual-rail case of classical communication.}  \label{fig:dualMI}
  \end{center}
\end{figure}

\section{Achievable communication rates}
\label{sec:channelcapacities}

We start from the states \eqref{eq:psiS}, \eqref{eq:psiD}, \eqref{eq:state2}, and \eqref{eq:state3}; then we transform the part of the state watched by the accelerated observers to the Rindler basis.
The transformations between the Unruh and Rindler bases are taken directly from \cite{Bruschi2010}.
This density matrix is found numerically, using a truncated Fock basis for Rob and anti-Rob.
We do this since beyond the single mode approximation ($q_\text{R}<1$) the relevant density matrices are no longer block diagonal and they cannot be diagonalized into a closed form, as can be seen in  \cite{Bruschi2010}.

We then perform the partial trace over the observer not involved in the particular communication we are calculating.
We also perform the partial traces over each party in the bipartition.
Finally, we diagonalize these density matrices and use Eqs.~\eqref{eq:holevoinfo} and~\eqref{eq:CEdefinition}.

We compute the conditional entropy using states \eqref{eq:psiS} and \eqref{eq:psiD} for the two quantum channels, channel 1 where Alice sends information to Rob and channel 2 where Alice sends information to anti-Rob.
We find that if one of the channels is able to generate entanglement, the other cannot.

More precisely, the sum of the conditional entropies of both channels is always greater than or equal to zero.
This is a consequence of the strong subadditivity of the von Neumann entropy \cite{Lieb1973}, which implies that for any tripartite system composed of parties $A$, $B$, and $C$, the inequality,
\begin{align}
S(\rho_A)+S(\rho_C)\leq S(\rho_{AB})+S(\rho_{BC}),
\end{align}
holds.
Choosing $A=\text{anti-Rob}$, $B=\text{Alice}$, and $C=\text{Rob}$ and rearranging, we obtain this condition on the sum of conditional entropies,
\begin{align}
[S(\rho_{A\bar{R}})-S(\rho_{\bar{R}})]+[S(\rho_{AR})-S(\rho_R)]\geq 0.
\label{sumcond}
\end{align}
If Alice can generate quantum entanglement with Rob, she will not be able to do the same with anti-Rob, and vice versa.
The results we are presenting saturate the inequality \eqref{sumcond}, but the interpretation is the same: the entanglement generation ability of the state merging protocol is bound to be zero for at least one of the bipartitions. Note that this result is valid for any tripartite quantum system, and thus it is not a specific feature of the relativistic setting we are considering. Although this inequality imposes a restriction to quantum communication if we insist in keeping a fixed Unruh mode, it is no barrier to quantum communication with both receivers for a broadcast channel, as shown in \cite{Dupuis2011}.

\begin{figure}[bt]
  \begin{center}
      \includegraphics[width=85mm]{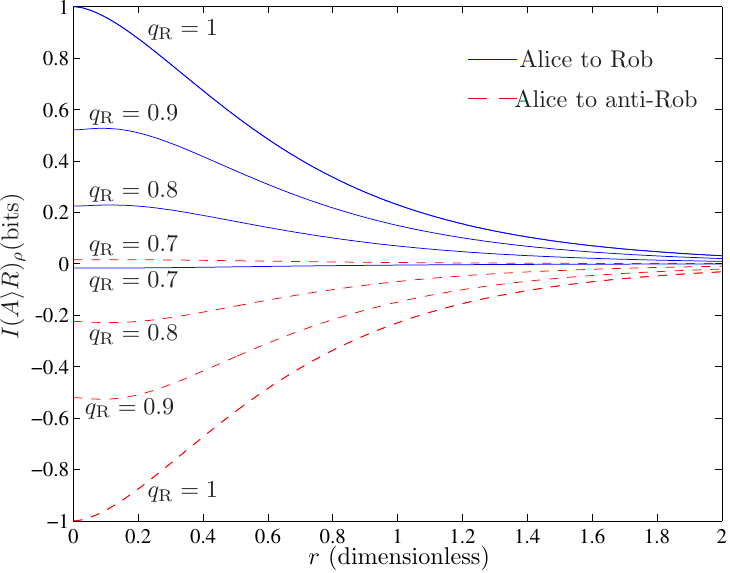}
  \caption[Single rail coherent information]{(Color online) Coherent information of the single-rail case of quantum entanglement generation.}  \label{fig:singleNCE}
  \end{center}
\end{figure}

\begin{figure}[tb]
  \begin{center}
      \includegraphics[width=85mm]{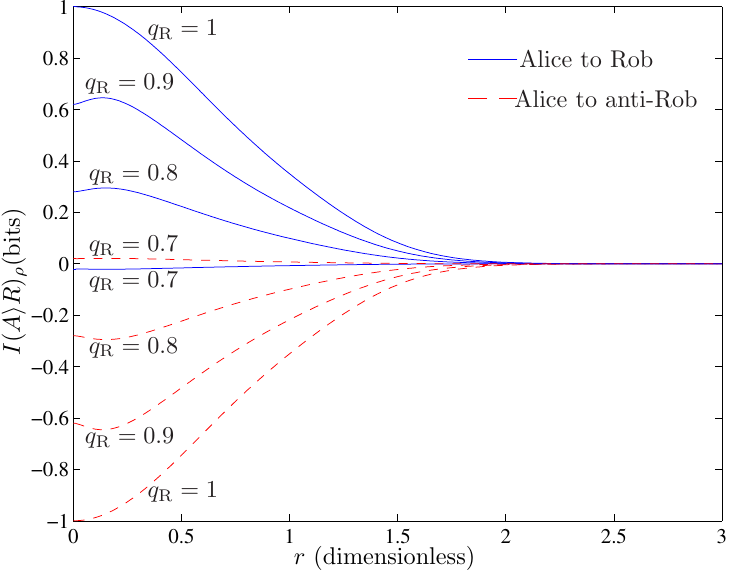}
  \caption[Dual rail coherent information]{(Color online) Coherent information of the dual-rail case of quantum entanglement generation.}  \label{fig:dualNCE}
  \end{center}
\end{figure}

\section{Results}
\label{sec:results}
The achievable rate of classical communication is given by the Holevo information maximized over the input parameters.
This has been calculated and plotted for both the single-rail case in Fig.~\ref{fig:singleMI} and the dual-rail case in Fig.~\ref{fig:dualMI}.

We see from the plots that for large acceleration the maximum Holevo information is not achieved for $\qr=1$ (single mode approximation).
This shows that the assumptions made in \cite{Hosler2012} are not correct for larger accelerations.
While it is true that SMA maximizes communication for low accelerations, for large accelerations we need to go beyond the SMA to achieve the maximum.

Classical communication tends to a finite value for large accelerations (equivalent to large $r$), which is larger for dual-rail communication.
No matter what the value of $\qr$ we find that as acceleration increases it is possible for Alice to communicate with both Rob and anti-Rob.
At large accelerations Alice has equal communication channel capacity with both Rob and anti-Rob, which is optimized at $\qr=\sqrt{0.5}$.

Note from Figs.~\ref{fig:singleMI}~and~\ref{fig:dualMI} that, at infinite acceleration, the Holevo information of the Alice-Rob and Alice--anti-Rob channels both converge to the same value.
This is easily understood by looking at Eqs. \eq{umodes} and the definition of $C_{\omega,\text{L}}$ and $C_{\omega,\text{R}}$, and noting that at infinite acceleration both $\sinh\ro$ and $\cosh\ro$ tend to $e^{\ro}/2$, and therefore in this limit \eq{umodes}  may be written as
\begin{align}
C_{\omega,\text{U}}\approx \frac{e^{\ro}}{2}\left[\qr a_{\text{I},\omega}-\qr a^\dagger_{\text{II},\omega}+q_\text{L} a_{\text{II},\omega}-q_\text{L}a^\dagger_{\text{I},\omega}\right].
\end{align}
This expression is invariant under the replacement $\text{I}\leftrightarrow\text{II}$, $q_\text{L}\leftrightarrow q_\text{R}$, which takes Rob to anti-Rob and vice versa, and therefore in the infinite acceleration limit the Holevo information of both bipartitions must be the same.

The quantum coherent information is given by the negative of the conditional entropy.
This has been calculated and plotted for both the single-rail in Fig.~\ref{fig:singleNCE} and the dual-rail in Fig.~\ref{fig:dualNCE}.

The plots show that when the coherent information is positive for one bipartition, it is negative for the other.
We call this the monogamy property: Alice must choose in advance, when choosing the fixed Unruh mode she is going to use, with whom she wants to generate entanglement for quantum communication. This is not an issue when one considers the full formalism of quantum broadcast channels; see \cite{Dupuis2011}.
For large acceleration, the coherent information tends to zero, for both single-rail and dual-rail methods.
The dual-rail methods always perform slightly better than the single-rail methods.

For both classical and quantum communication, we find that beyond the SMA the decrease in information transfer is nonmonotonic.
However, in most cases this is not optimal, as Alice is able to choose her modes, and therefore has control over the value of $\qr$.
This means that, after the maximization, the communication rates are monotonically decreasing with the acceleration parameter.

\begin{figure}[tb]
  \begin{center}
      \includegraphics[width=85mm]{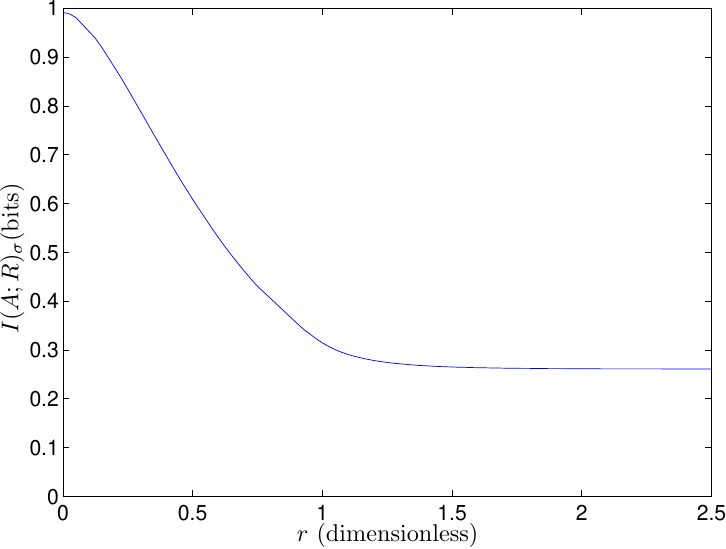}
  \caption[Single rail optimal MI]{(Color online) Optimal Holevo information as a function of the acceleration in the single-rail case.}  \label{fig:opt}
  \end{center}
\end{figure}

\begin{figure}[tb]
  \begin{center}
      \includegraphics[width=85mm]{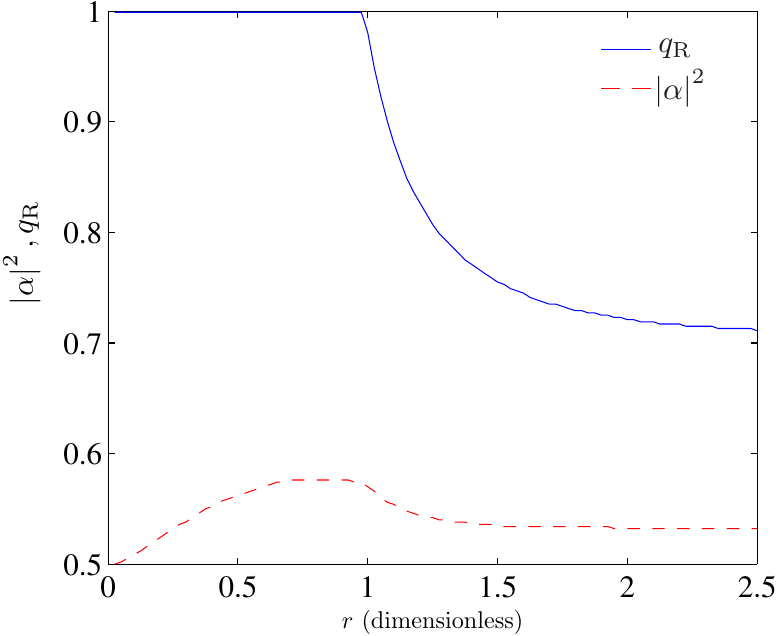}
  \caption[Single rail optimal parameters MI]{(Color online) $\modsq{\alpha}$ and $\qr$ optimal parameters for the results plotted in Fig. \ref{fig:opt}.}  \label{fig:opt2}
  \end{center}
\end{figure}

We maximized coherent information as a function of both $\modsq{\alpha}$ [in \eq{eq:psiD}] and $\qr$ [in \eq{umodes}], as these are the two parameters which can be controlled by Alice.
For the single-rail case, we found that $\qr=1$ is always optimal, in accordance with the assumption made in previous work \cite{Hosler2012}, whereas the optimal $\modsq{\alpha}$ is always close to $\nicefrac{1}{2}$, but not exactly equal.
As for the dual-rail case, due to the symmetry between the excitations, the value $\modsq{\alpha}=\nicefrac{1}{2}$ is always optimal, as is $\qr=1$.

The same computations are more contrived in the classical case.
From Figs.~\ref{fig:singleMI}~and~\ref{fig:dualMI} we see that the SMA is not optimal for all accelerations, so we expect a more varied behaviour of the optimal $(\modsq{\alpha},\qr)$ pair.
Fig.~\ref{fig:opt} shows the optimal Holevo information as a function of $\ro$ for this optimal pair of values, both for single- and dual-rail cases.
Fig.~\ref{fig:opt2} shows the optimal values of $(\modsq{\alpha},\qr)$.
We see that the single mode approximation is optimal up to $\ro\approx1$.
Note the interesting nonmonotonic behaviour of the parameter $\modsq{\alpha}$, which starts at approximately the same time $\qr=1$ becomes nonoptimal.
This parameter is always close to $\nicefrac{1}{2}$, but slightly biased to higher values.
This is due to the asymmetry in the noise when using the single-rail method.

\section{Conclusion}
\label{sec:conclusion}
We have studied the classical Holevo information and the quantum conditional information (both in the single- and dual-rail scenarios) in the situation where an inertial observer (Alice) communicates by sending information to two counteraccelerating observers each outside of the other's acceleration horizon (Rob and anti-Rob).

We found that the quantum channels between the inertial observer and each of the accelerated observers are mutually exclusive, by which we mean that for any particular choice of Unruh modes, the mutual information between Alice and one of the accelerated observers always vanishes. Although this provides a constraint to quantum communication in this restricted setting, it is not an issue when considering the full quantum broadcast channel, as shown in \cite{Dupuis2011}.

We showed that an Unruh mode with $\qr=1$ is always optimal to send quantum information to Rob and $\qr=0$ is optimal for communication with anti-Rob.
This is related to the subadditivity of the quantum channel capacities.
We find that for our setting the inequality saturates.

The classical channels are not mutually exclusive, so Alice is able to send classical information to both Rob and anti-Rob simultaneously.

For larger acceleration, and therefore larger $r$, we find that $\qr=1$ is no longer optimal for sending classical information to Rob, in both the single- and dual-rail methods.
We have computed the optimal $\modsq{\alpha}$ and $\qr$, showing that the corresponding Holevo information is a monotonically decreasing function of the acceleration.

\section{Acknowledgements}
The authors would like to thank Pieter Kok for useful discussions and Mark Wilde for helpful suggestions.
D.H. would like to thank, IFF, CSIC, Madrid for their hospitality.
D.H. was funded in part by a Santander Research Mobility Award.
D.H. was funded in part by a University of Sheffield Scholarship.

\end{document}